\documentstyle[aps,prl,epsfig]{revtex}

\begin{document}
\title{
 Infinitesimal incommensurate stripe phase in an axial
 next-nearest-neighbor Ising model in two dimensions
}
\author{Takashi Shirahata and Tota Nakamura}
\address {Department of Applied Physics, Tohoku University, Sendai, Miyagi
  980-8579, Japan}
 \date{\today}
\maketitle

\begin{abstract}
An axial next-nearest-neighbor Ising (ANNNI) model
is studied by using the non-equilibrium relaxation method. 
We find that the incommensurate stripe phase between the ordered phase
and the paramagnetic phase 
is negligibly narrow or may vanish
in the thermodynamic limit.
The phase transition is the second-order transition 
if approached from the ordered phase,
and it is of the Kosterlitz-Thouless type if approached from the
paramagnetic phase.
Both transition temperatures 
coincide with each other within the numerical errors.
The incommensurate phase which has been observed previously is a
paramagnetic phase with a very long correlation length
(typically $\xi\ge 500$).
We could resolve this phase by treating very large systems 
($\sim 6400\times 6400$), which is first made possible by employing
the present method.
\end{abstract}


\twocolumn

\section{INTRODUCTION}
\label{sec1}
Of late years, incommensurate (IC) stripe structures 
have been interesting subjects in various physical phenomena.
As the typical examples, we may list
an alloy ${\rm Er_{90}Y_{10-x}La_{x}}$ which contains heavy
rare earth metals, {\rm Er, Tm},\cite{Kawano,Elliott}
the incommensurate phase of dielectric material such as
${\rm Pb(Zr_{1-x}Ti_x)O_3}$ and
${\rm NaNO_2}$,\cite{Dai,Ricote,Viehland,Watanabe,Massidda}
and the stripe structure in ${\rm CuO_2}$
planes of oxide superconductors.\cite{Tranquada}
In ${\rm Er_{90}Y_{10-x}La_{x}}$, 
the longitudinal incommensurate oscillatory phase appears
between the paramagnetic phase and the ordered phase.
The aligned holes (domain walls) separate antiferromagnetic
stripes in ${\rm CuO_2}$ planes of oxide superconductors, then the spin
and charge are modulated.
In such systems, cooperative effects of
fluctuation and frustration are considered to play important roles.
Thus, we sometimes treat them with the axial next-nearest-neighbor Ising
(ANNNI) model as the simplified theoretical model.
For instance, when a uniaxial anisotropy is strong in the dielectric 
material, the Hamiltonian is equivalent to the
ANNNI model if we only consider the dipole interactions up to
the next-nearest-neighbor distance.
The phase diagram of ${\rm Pb(Zr_{1-x}Ti_x)O_3}$
obtained by experiments\cite{Dai}
agrees with that of the three-dimensional ANNNI model obtained by the
mean-field approximation.\cite{Bak}
In copper oxide materials, ${\rm Ca_{8}La_{6}Cu_{24}O_{41}}$ and
${\rm Ca_{2}Y_{2}Cu_{5}O_{10}}$,  
the Cu-O-Cu chains with ferromagnetic nearest-neighbor and
antiferromagnetic next-nearest-neighbor interactions are aligned
on two-dimensional planes whose interchain interactions are  
antiferromagnetic.
Furthermore, the spins on this plane are predicted to have 
a strong Ising anisotropy.\cite{Matsuda,Fong}
Consequently, we may treat these copper oxide planes as the
two-dimensional ANNNI model, between which the conduction electron
planes exist.

In the ANNNI model, there are exchange interactions up to the
next-nearest-neighbor pairs along one axis,
while there are only the nearest-neighbor interactions along the other
axes.
Most commonly, we take a convention that
the nearest-neighbor interactions are ferromagnetic and the 
next-nearest-neighbor ones are antiferromagnetic,
which cause frustration.
When the next-nearest-neighbor interactions (frustration) are weak, a
ferromagnetic state is the ground state, and a
paramagnetic-ferromagnetic phase transition occurs.
On the other hand, when frustration is strong, 
the ground state is the {\it antiphase}
state (abbreviated by $\langle 2 \rangle$), which is a commensurate(C)
stripe structure
of two up-spins and two down-spins like
$\uparrow\uparrow\downarrow\downarrow\uparrow\uparrow$.
It is widely accepted that the incommensurate stripe phase exists
between the paramagnetic phase and the antiphase, 
and that the successive phase transitions 
(paramagnetic-IC-antiphase) take place;
there is a ``finite'' incommensurate phase 
where the up-spins and down-spins are
aligned with a period longer than two.

Although only a topology of the phase diagram is known by the 
mean-field approximation in the three dimensional model,\cite{Bak}
estimates for the phase transition
temperatures in two dimensions have been done by several 
approximation theories
\cite{Muller,Villain,Kroemer,Grynberg,Saqi,Murai} and
numerical simulations.\cite{Selke1,Selke2,Sato}
However, values of the phase transition temperatures
scatter much depending on the methods employed.
One of the reasons that make the estimate difficult is a lack of
a reliable numerical simulation.
Because of frustration, the correlation time of Monte Carlo (MC)
simulations becomes very long near the critical point in large systems.
Thus, we are hardly able to reach thermal equilibrium
states within reasonable time steps.
A powerful method called a cluster heat bath (CHB) method\cite{Sato}
was developed to reduce the
correlation time, however, the system size accessible
by this method is restricted up to about $64\times 128$ lattice sites.

Recently, a new method using a MC simulation has been developed: the 
non-equilibrium relaxation (NER) method.\cite{Ito0,Ito1}
We are able to understand phase transitions from the differences of
behaviors in non-equilibrium relaxation processes, which have
been discarded in conventional MC simulations.
Since we do not wait until the equilibrium is realized, we can use the 
CPU time to enlarge the system sizes.
Therefore, it makes possible to treat large systems that can not be
possible by other methods.
Accordingly, there expected to be little finite-size effect in the
obtained data, and thus
we can regard them as those of the infinite systems.
By reasons stated above,
the NER method can be effective especially in systems with a slow
dynamics which a very long equilibration is necessary.
\cite{Ito2,Ito3,Ozeki1,Ito4,Ogawa,Nonomura1,Nonomura2,Ozeki2,Ozeki3}

In this paper, we study the two-dimensional ANNNI model by the
NER method in order to determine the successive phase transition
temperatures.
However, the obtained results suggest that the IC phase may disappear
within limits of numerical errors.
Moreover, we found a very exotic phase transition,
which seems to be the Kosterlitz-Thouless (KT) transition\cite{KT} 
if it is observed from the high temperature side,
and seems to be the second-order transition if observed from the low
temperature side.
This evidence suggests that the frustration parameter does not serve
as an asymmetric parameter\cite{Schulz,Selke3,Ostlund}
which explicitly favors the IC structure.

We describe the model Hamiltonian and the NER method in Sec.~\ref{sec2},
and present simulational results in Sec.~\ref{sec3}.
Section~\ref{sec4} is devoted to conclusions.

\section{MODEL AND METHOD}
\label{sec2}
\subsection{Two-dimensional ANNNI model}
\label{sec2_0}
The two-dimensional ANNNI model is described by the following Hamiltonian,
\begin{eqnarray}
 {\cal H} & = & -\sum_{x,y}(J_0 S_{x,y}S_{x+1,y}+J_1 S_{x,y}S_{x,y+1}
 \nonumber\\
 &&+J_2 S_{x,y}S_{x,y+2}), 
\end{eqnarray}
where $J_0(>0)$ is the nearest-neighbor interaction along 
the chain direction, which is the direction that has no frustration.
$J_1(>0)$ and $J_2(<0)$ are respectively the nearest-neighbor
and the next-nearest-neighbor interactions along a direction
perpendicular to the chain direction
(axial direction that has frustration),
and $S_{x,y}=\pm 1$.
In this paper, we fix $J_0=J_1$ for simplicity,
and impose the open boundary conditions along the axial ($y$) direction,
while we use the periodic boundary conditions along the chain ($x$)
direction.
We define a ratio between the nearest and the next-nearest-neighbor
interactions along the axial direction as $\kappa(=-J_2/J_1)$.
Our interest is restricted to the region $\kappa>1/2$, where the
successive phase transitions 
(${\rm antiphase}\rightarrow {\rm IC}\rightarrow {\rm paramagnetic}$)
has been considered to occur.

The fermion approximations assumed that the system has straight
domain walls along the chain direction at the intermediate 
temperatures.\cite{Villain}
That is, domains whose periods are
longer than two
appear among the commensurate antiphase domains. 
In this case, the spin structure becomes incommensurate along the axial
direction, while the spins along the chain direction order 
ferromagnetically.
On the other hand, it was postulated that domain walls run along 
the axial direction in the
interface free energy method of ${\rm M\ddot{u}ller}$-Hartmann and
Zittartz(MHZ).\cite{Muller,Kroemer}
Namely, the correlation of spins ordered ferromagnetically
along the chain direction is destroyed by these domain walls.

Sato and Matsubara\cite{Sato} discussed using the CHB
simulation that the transition temperature ($T_{c2}$)
between the antiphase and the IC phase agrees with that
of the free-fermion approximation while the transition
temperature ($T_{c1}$) between the IC phase and the
paramagnetic phase is close to $T_{c2}$ obtained by MHZ.
Thus, they considered that the domain walls penetrate into the system
along the chain direction at $T_{c2}$
as the temperature increases from the ground state.
The ferromagnetic correlations along the
chain direction remain until they are destroyed at $T_{c1}$ and
the paramagnetic state is realized.
This idea means that the IC phase exists between the penetration of
incommensurate domains in the {\it axial} direction and 
the disappearance of the ferromagnetic {\it chain} correlation.
Therefore, the spin structure along the axial direction is
incommensurate, while it is ferromagnetic along the chain direction 
in the IC phase.
It has been considered that the phase transition between the paramagnetic 
phase and the IC phase is the KT transition and that between
the IC phase and the antiphase is the second-order transition.

We consider following two quantities to clarify the phase transitions.
One is the {\it antiphase magnetization} defined by
\begin{eqnarray}
 m_{\langle 2 \rangle}(t)
 =\frac{1}{N}\sum_{x,y}S_{x,y}(\langle 2 \rangle)S_{x,y}(t),
 \label{anti_mag}
\end{eqnarray}
where $N$ is the total number of spins in the system,
$S_{x,y}(t)$ denotes the spin value of the site $(x,y)$ 
at time $t$, and $S_{x,y}(\langle 2 \rangle)$ represents
the antiphase ordered state.
The antiphase magnetization takes a finite value in the antiphase, but
vanishes in the IC phase and the paramagnetic phase, since the domain
walls destroy the antiphase state at the lower
transition temperature, $T_{c2}$.
Therefore, this parameter is employed to estimate $T_{c2}$.

Here, we should note that 
the antiphase magnetization is not a relevant order parameter
that decays algebraically at the transition temperature.
Since the elementary excitation is a domain wall which readily
percolates the chain direction, the antiphase magnetization
decays exponentially once the domain wall penetrate into the system.
Usually, the density of the domain walls has been used 
as an order parameter.
This parameter is considered to diverge algebraically 
at the transition temperature as $(T-T_{c2})^\beta$,
which is equivalent to that the correlation length diverges
algebraically as $\xi\sim(T-T_{c2})^{-\nu}$.
In this point, the phase transition between the IC phase and
the antiphase is the second-order transition.
In the Monte Carlo simulations, the correlation length 
is related to the characteristic time as $\tau\sim\xi^{z}$,
where $z$ is called the dynamic exponent. 
Therefore, the characteristic time also diverges at $T_{c2}$
as $\tau\sim(T-T_{c2})^{-z\nu}$.
We can extract it
from the relaxation of the antiphase magnetization: 
the time that the antiphase magnetization begins to decay 
exponentially is the characteristic time that the domain wall 
penetrates into the system.
We execute the finite-time scaling to obtain the characteristic
time at each temperature
and estimate $T_{c2}$ as its diverging temperature.
A concrete procedure of the finite-time scaling
is explained in Sec.~\ref{sec2_3}.
The antiphase magnetization is an extensive variable
so that it shows better accuracy as the system
size is enlarged because of the self averaging.
Thus, we use the antiphase magnetization
to estimate the lower transition temperature, $T_{c2}$.

The other quantity is the {\it layer magnetization} defined as
\begin{eqnarray}
 m_{l}(t)=\frac{1}{L_y}\sum_{y=1}^{L_y}\left(\frac{1}{L_x}\sum_{x=1}^{L_x}
 S_{x,y}(t)\right)^2,
 \label{layer_mag}
\end{eqnarray}
where $L_x$ and $L_y$ are the length of the system along the chain and
the axial direction, respectively.
The layer magnetization reflects the spin order along the chain
direction.
In the IC phase, the incommensuration is realized in the axial
direction.
The spin arrangement along the chain direction should order 
ferromagnetically or at least be critical.
The free energy of this state, which is of the order of 
$-L^{1+\frac{1}{\nu}}$, is always lower than that of the state
with the spins disordered only along the chain direction, which is 
of the order of $-L$, if $\nu$ is positive.
Accordingly, the layer magnetization vanishes exponentially 
in the paramagnetic phase,
but takes a finite value or
behaves in a power law in the IC phase.
Therefore, this order parameter is used to estimate the upper transition
temperature, $T_{c1}$.
We determine the transition temperatures by using these two quantities
in this paper.

\subsection{Non-equilibrium relaxation (NER) method}
\label{sec2_1}
Phase transitions occur in the
infinite-size limit($L\rightarrow\infty$) 
and in the equilibrium limit($t\rightarrow\infty$).
Because we cannot take both limits at the same time in the simulation,
the equilibrium limit has been taken first conventionally, and then, 
we take the infinite-size limit by using the finite-size scaling.
However, the dynamics of the simulations are very slow in the 
frustrated systems, which causes a very large correlation time.
From a time-space relation, $\tau\sim\xi^{z}$, this means that the 
correlation length is also very large.
As will be mentioned in Sec.~\ref{sec3},
we estimate the correlation length, $\xi_x\sim500$, for $\kappa=0.8$
and $T=1.40$, which resides in the IC phase of previous phase diagram.
In this situation, reliability of the finite-size scaling might become
doubtful.
In this paper, we follow a completely alternative approach to the
thermodynamic limit, i.e.,
we observe the relaxation of the infinite-size system to the equilibrium
state.
In order to extract the equilibrium properties, the finite-time scaling
analysis is utilized instead of the conventional finite-size scaling.
This approach is known as the non-equilibrium relaxation (NER) method.
\cite{Ito0,Ito1,Ozeki2,Ozeki3}
Actually, we prepare a very large lattice 
and observe the relaxation of physical quantities.
The simulations are stopped before the finite-size effect appears.
Accordingly we can regard the systems as the infinite systems.

Using the NER method, we can estimate the phase transition temperature
and the critical exponents by examining behaviors of the relaxation
processes to the thermal equilibrium state
(non-equilibrium relaxation processes).
The analysis is based on the dynamic finite-size scaling hypothesis of
the free energy that
\begin{eqnarray}
 F(\varepsilon,h,L,t)=L^{-d}\tilde{F}
 (\varepsilon L^{\frac{1}{\nu}},hL^{d-\frac{\beta}{\nu}},tL^{-z}),
 \label{dfss_free}
\end{eqnarray}
where $\varepsilon(\equiv(T-T_c)/T_c),h,L,t$ are the relative temperature,
the symmetry breaking field, the system size and time, respectively.
$\nu$ and $\beta$ denote static exponents, while $z$ is a dynamic
exponent, and $d$ is a dimension of space.\cite{Suzuki}
The order parameter is given by a derivative of the free energy with the
field:
\begin{eqnarray}
 m(t)\sim \left.\frac{\partial F}{\partial h}\right|_{h=0}
 &=&L^{-\frac{\beta}{\nu}}\bar{F}(\varepsilon L^{\frac{1}{\nu}},tL^{-z})
 \nonumber\\
 &=&t^{-\frac{\beta}{z\nu}}\hat{F}(\varepsilon t^{\frac{1}{z\nu}})\nonumber\\
 &\propto& t^{-\frac{\beta}{z\nu}}(\varepsilon=0,L\rightarrow \infty),
 \label{dfss_ord}
\end{eqnarray}
where we have set $L\sim t^{1/z}$ at the transition point
($\varepsilon=0$) because the characteristic time scale $\tau$ and 
the correlation length $\xi$ should scale as
$\xi\sim\tau^{1/z}$.
Equation~(\ref{dfss_ord}) describes that the order parameter decays in a
power law with time at the transition temperature.
On the other hand, the order parameter behaves exponentially at 
temperatures away from the critical point.
Thus, we are able to estimate $T_c$ by this
difference.
Actually, we measure the order parameter $m(t)$ at each MC step $(t)$
started from an ordered spin configuration at a given temperature.
We repeat this MC run by changing the random number seeds, and $m(t)$ is
averaged over these independent MC runs.
The temperature at which the $m(t)$
curve decays in a power law is the transition temperature.

We use the local exponent to ascertain whether the order parameter
decays in a power law or exponentially,
which is defined by
\begin{equation}
 \lambda(t)=\left|\frac{d\log m(t)}{d\log t}\right| .
\end{equation}
When we plot the local exponent against $1/t$, it diverges to infinity for
$T>T_c$, it converges to zero for $T<T_c$ and to a finite value
($\ne 0$) at $T=T_c$ in the limit of $1/t\rightarrow 0$.
The upper bound of $T_c$ is the lowest temperature that $\lambda(t)$
diverges, and the lower bound is the highest temperature that 
$\lambda(t)$ decays to zero.
The convergent value of the local exponent at $T=T_c$ is the
critical exponent $\lambda=\beta/z\nu$ from Eq.~(\ref{dfss_ord}).

\subsection{NER of fluctuation}
\label{sec2_2}
Here, we describe the NER of fluctuation.
\cite{Ito1,Ito2,Jaster,Sadiq}
The susceptibility is written by differentiating 
Eq.~(\ref{dfss_free}) with the symmetry breaking field twice,
\begin{eqnarray}
 \left.\frac{\partial^2 F}{\partial h^2}\right|_{h=0}
 &\propto&\langle m(t)^2\rangle -\langle m(t)\rangle^2\nonumber\\
 &\propto& t^{\frac{d}{z}-\frac{2\beta}{z\nu}}=t^{\frac{\gamma}{z\nu}},
 \label{dfss_chi}
\end{eqnarray}
where we have used the scaling relations:
\begin{eqnarray}
 \cases{
  d\nu=2-\alpha \cr
  \alpha+2\beta+\gamma=2. \cr
 }
\end{eqnarray}
The susceptibility diverges in a power law at the transition
temperature.
Thus, we are able to estimate the transition temperature and the
critical exponents from the susceptibility.
It is also noticed that the NER of fluctuation does not require us to
start with a symmetry-broken ordered state.
The quantity of fluctuation always takes a definite value and diverges
at $T_c$ even though the symmetry is not spontaneously broken.
Therefore, we can start from a paramagnetic state or any state in this
scheme.
This is especially useful when an ordered state is not known yet,
or it is difficult to realize.
The layer magnetization defined by Eq.~(\ref{layer_mag}) is 
equivalent to the second 
derivative with respect to a local field along a single chain
in the paramagnetic phase.
When the simulation starts from the antiphase, the first derivative
term remains finite in the NER process even though the temperature 
is in the paramagnetic phase.
Therefore, the layer magnetization is not regarded as 
the susceptibility.
The NER of the layer magnetization is within a scheme of 
the NER of fluctuation, only when the simulation is started
from the paramagnetic state.

\subsection{Finite-time scaling}
\label{sec2_3}
In case when the NER function does not begin to decay algebraically 
within a reasonable time, it is difficult or almost impossible
to estimate the transition temperature directly by the local exponent.
Even in such a situation, we are able to determine it by using the
finite-time scaling analysis, which is a direct interpretation of the
finite-``size'' scaling by a relation $\xi\sim\tau^{1/z}$.
We present the finite-time scaling relation \cite{Ozeki2,Ozeki3}
as follows,
\begin{equation}
 m(\varepsilon,t)=t^{-\lambda}\hat{m}(t/\tau(\varepsilon)),
 \label{ft_scaling}
\end{equation}
where $\lambda=\beta/z\nu$, and $\tau(\varepsilon)$ denotes a relaxation
time at the relative temperature $\varepsilon$.
Since the relaxation time diverges algebraically 
in the case of the second-order phase transitions, 
the relation between the relaxation time and the relative temperature is
described by
\begin{equation}
 \tau(\varepsilon)=A\varepsilon^{-z\nu}.
 \label{2_trans_sc}
\end{equation}
On the other hand, in the case of the KT transition,\cite{KT}
it is considered that the relaxation time diverges exponentially, which
we assume that
\begin{equation}
 \tau(\varepsilon)=A\exp(B/\sqrt{\varepsilon}).
 \label{KT_trans_sc}
\end{equation}

Now, we describe how we actually estimate the transition temperature by
using Eq.~(\ref{ft_scaling})$\sim$(\ref{KT_trans_sc}).
We use only data which is clearly in the paramagnetic phase; $T>T_c$.
First, we plot $m(t)t^\lambda$ at various temperature against
$t/\tau(\varepsilon)$ by using Eq.~(\ref{ft_scaling})
to determine $\lambda$ and $\tau(\varepsilon)$ so that
all data points fall on a single curve.
Next, we plot $\tau(\varepsilon)$ against $\varepsilon$, and
fit the points to a smooth curve as we change $T_c$ and $z\nu$
by using Eq.~(\ref{2_trans_sc}) for the second-order transition. 
In the KT-transition case, we use Eq.~(\ref{KT_trans_sc}) instead.
The temperature at which the least-square fitting error becomes 
minimum is the most probable estimate for $T_c$.
The phase transition between the paramagnetic
phase and the IC phase is the KT transition and 
that between the IC phase and the antiphase is the
second-order transition in the two-dimensional ANNNI model.
Therefore, we use Eq.~(\ref{KT_trans_sc}) for the upper transition
temperature, and Eq.~(\ref{2_trans_sc}) for the lower one.

\section{SIMULATION AND RESULTS}
\label{sec3}
\subsection{NER from the antiphase state}
We examine the phase transition temperature for $\kappa=0.6$
and $0.8$.
The temperature is measured in a unit of $J_1$.
Here, the NER of two quantities are presented by the 
simulation started from the antiphase state.

First, we observe the NER of the
layer magnetization 
to check the relation between the finite size, $L_x$,
and the correlation length, $\xi_x$, along the chain direction. 
We change the size along the chain
direction from $L_x=799$ to $L_x=25599$, while the length along the
axial direction is fixed to $L_y=800$.
Figure1 shows the NER of the layer magnetization,
Eq.~(\ref{layer_mag}), for $\kappa=0.8$ and $T=1.40$.
This temperature belongs to the IC phase (KT phase)
in previous investigations as summarized in Table~\ref{Tc_comparison}.
When the system size is small ($N=799\times 800$), the relaxation
roughly looks like a power-law decay, which misled us to 
the KT phase.
As the system becomes larger, the relaxation exhibits an exponential
decay confirming us that the system is in the paramagnetic phase.
Note that the convergence to a finite value is due to the finite-size
effect.
By its definition, the equilibrium value of the layer magnetization
is roughly estimated as $\xi_{x}/L_x$ in the paramagnetic phase.
As shown in Fig.1, the convergence value takes a half-value
if the system size is doubled.
Thus, we can estimate the correlation length $\xi_x\sim 500$
in this system.
From this figure, we can also understand the
relation between the finite size-effect and the time-effect; 
the effective time that the system behaves as the infinite size.
For example, it is about $15000$ MCS
in the system with $L_x=6399$.
After this time scale, finite size effect appears in dynamics.
In this subsection, we use the lattice of $N=6399\times 6400$
and determine the observing time at each temperature
until which the relaxation curve do not begin to bend 
to an equilibrium value.
We average $24\sim32$ independent MC runs at each temperature.

Figure2 is a raw data of the NER of the antiphase
magnetization, Eq.~(\ref{anti_mag}), for $\kappa=0.6$.
At all the temperatures, the antiphase magnetization clearly decays
exponentially, which guarantees that the transition temperature
must be lower than 0.98.
As mentioned in Sec.~\ref{sec2_0},
the antiphase magnetization is not a relevant
order parameter, and so it decays exponentially as soon as
domain walls penetrate into the system.
Therefore, we estimate the transition temperature by using the
finite-time scaling.
At first, we determine the exponent, 
$\lambda$, and the relaxation time, 
$\tau_{\langle 2\rangle}(\varepsilon)$, at each temperature 
so that the scaled data, $m_{\langle 2\rangle}(t)t^{\lambda}$,
fall on a single scaling function when plotted against 
$t/\tau_{\langle 2\rangle}(\varepsilon)$ (Fig.3(a)).
Here, the relaxation times are normalized by the value of $T=1.06$,
and are listed in Table~\ref{tau_0.6}.
The exponent, $\lambda$, that gives best scaling is 
$\lambda=0.015$.
Excellence of the scaling shown in Fig.3
may yield the validity of the finite-time scaling hypothesis.

Next, we estimate $T_{c2}$ by Eq.~(\ref{2_trans_sc}) because it is
predicted that the phase transition between the antiphase
and the IC phase is the second-order.
We plot $\tau_{\langle 2\rangle}(\varepsilon)$ against 
$\varepsilon(=\frac{T-T_{c2}}{T_{c2}})$ as changing $T_{c2}$
and find a $T_{c2}$ that gives the best linearity in the log-log scale
as shown in Fig.3(b).
We obtain $T_{c2}=0.89\pm 0.02$ and $z\nu=4.07$.
The error $\pm 0.02$ is a range of temperature 
in which the data
points clearly fall on the fitting line.
The exponent has a range within which we
can perform a good scaling as $\lambda=0.000\sim0.030$.
The transition temperature takes the same value
irrespective of the choice of $\lambda$.
Evidences such as
the algebraic divergence of $\tau_{\langle 2\rangle}(\varepsilon)$
and excellence of the finite-time scaling
support that this transition is the second-order.
This is a clear distinction from the three dimensional model, 
where the lower transition is considered as the first-order.

We estimate the upper transition temperature, $T_{c1}$,
 using the layer magnetization,
Eq.~(\ref{layer_mag}), in the same way as mentioned above.
Figure4 shows raw data of the NER of the layer
magnetization, $m_{l}(t)$, for $\kappa=0.6$.
Here, the layer magnetization clearly decays exponentially,
and thus the $T_{c2}$ must be less than 0.98.
The finite-time scaling is shown in Fig.5(a).
The obtained relaxation time, $\tau_{l}(\varepsilon)$, is also
presented in Table.~\ref{tau_0.6}.
Since it is predicted that the phase transition between the IC phase
and the paramagnetic phase is the KT transition, we fitted 
$\tau_{l}(\varepsilon)$ by Eq.~(\ref{KT_trans_sc})
in Fig.5(b),
by which we obtain $T_{c1}=0.89\pm 0.02$ and $B=3.44$.
If we assumed that the phase transition is the second-order transition,
and performed the fitting by Eq.~(\ref{2_trans_sc}), the estimated
$T_{c1}$ becomes far from a physically meaningful value.
In consequence, we confirm that the upper phase transition is 
of the KT type.
Actually, Sato and Matsubara\cite{Sato} performed the finite-size
scaling of the layer magnetization supposing
\begin{equation}
 m_{l}\times L^{\eta}=Y[L^{-1}\exp(B\varepsilon^{-0.5})],
\end{equation}
which gave $T_{c1}=1.16$ and $\eta=0.25$ for $\kappa=0.6$.
This scaling form is equivalent to the finite-time scaling,
Eq.~(\ref{ft_scaling}) and (\ref{KT_trans_sc}), 
if we admit $L=t^{1/z}$ and $\lambda=\eta/z$.
The difference of the obtained transition temperatures can be attributed
to the difference of the system sizes. 
We estimate $\lambda=0.015$ by minimizing the normalized
residual, however, this value does not correspond to $(2-\eta)/z$.
Because we start simulations from the antiphase state, the
layer magnetization contains contributions from the antiphase
magnetization, $\langle m_{\langle 2\rangle}\rangle^2$.

We have obtained the $T_{c2}$ and the $T_{c1}$ 
for $\kappa=0.8$ in the same way.
In the finite-time scaling plot, we used the data at five temperature
ranging $T=1.40\sim 1.48$,
and obtained $T_{c2}=1.32\pm 0.02$ and $T_{c1}=1.31\pm 0.02$.

The phase transition temperatures, $T_{c1}$ and $T_{c2}$,
coincide within limits of error both for $\kappa=0.6$ and $0.8$.
This means the IC phase may vanish in the thermodynamic limit.
Because this is a rather daring conclusion, we must confirm it from
another point of view.
Actually, it might be dangerous to estimate the transition temperature
between the paramagnetic phase and the IC phase started from the ground
state.

\subsection{NER from the paramagnetic state}

We start the simulation from the paramagnetic state
and observe the NER of fluctuation; the layer magnetization.
Since the observable is a quantity of fluctuation,
it is necessary to take much more sample averages
compared to the NER of antiphase magnetization.
Therefore, the system size is restricted to $N=1999\times 2000$
at most.
The NER function of fluctuation diverges algebraically 
at $T=T_c$, diverges exponentially at $T<T_c$ and
remains finite at $T>T_c$.
We find the phase transition temperature from these differences.
It is noticed that the NER function finally converges to a finite value
because the system is finite.
Therefore, we must be aware of the range of time not to observe the
finite-size effect.
This is a time scale that the correlation length reaches the finite
system size.
We compare the NER functions of two different sizes 
($1999\times 2000$ and $1599\times 1600$/$1599\times 
 1600$ and $999\times 1000$/$1599\times1600$ 
 and $799\times 800$ et al.) for every data point, and
estimate this crossover time until which
two curves fall on the same line and the finite-size effect
does not appear. 
An example of this comparison is shown in Fig.6.
The NER functions of the layer magnetization from the paramagnetic
phase, $m_l(t)$, are plotted for three sizes,
$N=1999\times 2000$, $N=1599\times 1600$ and $N=1199\times 1200$
at $T=0.92$ and $\kappa=0.6$. 
Three curves fall on the same line until $25000$ MCS.
After this crossover time, however, the curve of $N=1199\times 1200$
bends down from the other two curves, probably 
because the correlation length reaches $1200$ at this temperature.
So, we have to discard the data of $N=1199\times1200$ after $25000$ MCS.
Comparing two curves of $N=1999\times 2000$ and $N=1599\times 1600$,
we can use the data of $N=1599\times 1600$ until
at least $10^5$ MCS.
Table~\ref{size_effect} shows the crossover MCS at each temperature 
for $\kappa=0.6$ and $0.8$.
Thus, we use a system of a proper size for a
proper observing time at each temperature.
We take averages over $1000\sim 7000$ independent MC runs.

The NER of the layer magnetization, $m_l(t)$, from the paramagnetic state
are shown in Fig.7 for (a)$\kappa=0.6$ and 
(b)$\kappa=0.8$.
Figure8 shows the corresponding local exponents. 
In Fig.8(a), the exponent decreases
for $T\ge 0.92$ and diverges for $T\le 0.88$.
At $T=0.90$, it converges to a finite value.
In consequence, we predict that the phase transition temperature between
the paramagnetic phase and the IC phase is $T_{c1}=0.90\pm 0.02$.
For $\kappa=0.8$, we obtained $T_{c1}=1.325\pm0.025$ as shown in
Fig.8(b).

Furthermore, we also analyze the upper transition temperature, $T_{c1}$,
using the finite-time scaling 
as shown in Figs.9 and 10.
We show the scaling in Fig.9 for
(a)$\kappa=0.6$ and (b)$\kappa=0.8$.
Since the layer magnetization diverges algebraically as 
$t^{\frac{2-\eta}{z}}$ at the transition temperature,
we plot $m_l(t)t^{-\lambda}$ against $t/\tau_l(\varepsilon)$.
All the curves excellently fall on the same line,
by which we obtain the relaxation time as shown 
in Fig.10.
Here, we adopt $\lambda=(2-\eta)/z=0.49$ that 
is a value which the local exponent converges to
(see Fig.8).
If we admit the KT-criterion $\eta=1/4$, the dynamic exponent
is estimated as $z\sim 3.6$.
Next, we fit the relaxation time using
Eq.~(\ref{KT_trans_sc}).
Figure10 shows the best least-square fitting 
for (a)$\kappa=0.6$ and (b)$\kappa=0.8$.
We obtain the transition temperature $T_{c1}=0.890\pm 0.015$ for
$\kappa=0.6$ and $T_{c1}=1.300\pm 0.013$ for $\kappa=0.8$.
These transition temperatures are again consistent with the values
estimated from the local exponents of the NER from the paramagnetic
state, and those by the scaling from the ground state.

Therefore, we are able to conclude that $T_{c1}$ and $T_{c2}$ 
coincide with each other within limits of error in both cases of
$\kappa=0.6$ and $0.8$.
The phase transition temperature 
between the paramagnetic phase and the IC phase is 
$T_{c1}=0.895\pm 0.025$ and that between the antiphase and the IC phase 
is $T_{c2}=0.89\pm 0.02$
for $\kappa=0.6$, and $T_{c1}=1.32\pm 0.03$ 
and $T_{c2}=1.32\pm 0.02$ for $\kappa=0.8$.
Since these temperatures are very close to each other, it is suggested
that the IC phase does not exist or it is very narrow 
even if it exists.
We need to pay attention to a fact that Monte Carlo simulation
is not able to exclude an very tiny temperature region.

\section{CONCLUSIONS}
\label{sec4}
It has been considered that the successive phase transitions
with a finite IC phase take place for $\kappa >0.5$
in two-dimensional ANNNI model.
In this paper, we estimated the phase transition temperatures by
applying the NER method,
and found that the $T_{c1}$ is equal to $T_{c2}$ within limits of
errors.
This is a very exotic phase transition, which is the KT type if
approached from the high-temperature side, and is the second-order 
if approached from the low-temperature side.
Therefore, we speculate successive phase transitions with an 
{\it infinitesimal} IC phase may occur in this system.

In the studies of the C-IC transitions in two-dimensional systems,
it has been investigated the systems with an asymmetric parameter
which explicitly favors the IC structure.
\cite{Schulz,Selke3,Ostlund}
For a finite value of the parameter, there exists the finite
IC phase between the commensurate phase and the paramagnetic phase.
In the limit of vanishing the parameter,
the IC phase shrinks to the infinitesimal.
The frustration parameter, $\kappa$, of the ANNNI model
has been considered as the asymmetric parameter based on
the approximate theory
\cite{Muller,Villain,Kroemer,Grynberg,Saqi,Murai} 
valid only at low temperatures,
and the small-scaled Monte Carlo simulations.\cite{Selke1,Selke2,Sato}
However, what is actually favored by frustration is the creation of 
the domain wall.
Between the creation of the domain wall and the realization of the 
IC structure, there are many conditions to satisfy, which have been
supposed in the free-fermion approximation.\cite{Villain}
One of these is the spin correlation along the ferromagnetic chain.
In the two-dimensional ANNNI model, 
the non-frustrated direction is only one dimension,
and thus the spin correlation along the ferromagnetic chain can be
easily destroyed.
Therefore, we question regarding the frustration parameter
as the asymmetric parameter.
If we consider that these two are not related with each other
and the asymmetric parameter is zero in the present model,
the width of the IC phase becomes infinitesimal.
This is what we observed in this paper.

Here, we show the comparison of the obtained transition temperatures 
with the previous ones in Table~\ref{Tc_comparison}.
It is recognized that our $T_{c1}$ is lower than any other ones,
though $T_{c2}$ is consistent with each other.
This can be explained by the difference of the finite-size effect.
The phase transition between the IC phase and the 
paramagnetic phase is confirmed to be the KT transition while the one
between the antiphase and the IC phase is the second-order.
The correlation length diverges algebraically against the temperature 
in the latter, while it diverges exponentially in the former case.
If a system size is small compared with the correlation
length and the accuracy of the numerical data is not enough,
even the finite-size scaling analysis may mislead to a wrong $T_c$,
where the finite but very large correlation length reaches the finite
system size.
Therefore, it is very likely that the KT transition temperature obtained
previously is over-estimated.
We may rather easily obtain the phase transition temperature accurately
in the second-order transition, because the divergence is
algebraically.
In the two-dimensional ANNNI model, the correlation length 
along the chain direction is 
$\xi_x\sim 500$ at $T=1.40$ and $\kappa=0.8$ which is the temperature a
little higher than the KT transition temperature 
and the correlation length is far larger
than the system sizes ever simulated previously 
($N=64\times 128$).\cite{Sato}
We actually used data with system sizes, 
$N=599\times 600 \sim 6399\times 6400$
in this paper, and checked the finite-size effect and the range of
observing time in which the system behaves as an
infinite system at each temperature.
Thus, our data can be considered as closed to the thermodynamic value.
This is a reason why we could detect the KT transition accurately.

In this paper,
we have supposed that the ferromagnetic correlation along the chain
direction remains finite or at least it is critical in the IC phase.
Thus, the $T_{c1}$ we have obtained is the temperature at which the
ferromagnetic correlation becomes critical.
Above this temperature, each ferromagnetic chain exhibits paramagnetism,
even though the correlation length is very large near the $T_{c1}$.
The present results only exclude the finite IC phase of this type.
In this study, it is clarified that the domain walls penetrate
into the system along the chain direction at the lower transition
temperature, $T_{c2}$.
The spin configuration along the
chain direction changes drastically from the ferromagnetic ordered 
state to the paramagnetic disordered state at the same temperature
within limits of error.
We can neglect the IC phase where the ferromagnetic correlation 
has been expected to be critical.
Therefore, the domain walls mostly run straight along the chain
direction.
We consider this is why the naive free-fermion approximation 
of Villain and Bak\cite{Villain} gives the best quantitative
agreements with our estimate of $T_{c2}$, 
and is the best approximation.
The analyses for other regions of $\kappa$ and determination of the
critical exponents will be a task in the future.
The NER of the Binder parameter, the specific heat and the spin
correlation with high accuracy is necessary.\cite{Ito1}

As for the dielectrics, ${\rm Pb(Zr_{1-x}Ti_{x})O_3}$,
the phase diagram\cite{Dai} in the low concentration of {\rm Ti} is very
similar to that of the ANNNI model if we interpret the paraelectric,
the ferroelectric and the antiferroelectric phases as the 
paramagnetic, the ferromagnetic and the antiphase phases.
There are two controversial explanations for the existence of the
incommensurate phase observed in the experiment.
Ricote {\it et al}.\cite{Ricote} concluded that the appearance of the IC
phase is due to the surface effect by two experiments using the powder
neutron diffraction and a transmission electron microscope.
On the other hand, Viehland {\it et al}.\cite{Viehland} considered it a
bulk effect by directly observing the high resolution image of a
transmission electron microscope.
In addition Watanabe {\it et al}.\cite{Watanabe} also described that it
is not because of the surface
effect by examining the stability of the IC phase in
the bulk.
If this compound can be explained by the ANNNI model,
the existence of the IC phase is only possible in three dimensions.
Therefore, we predict that the appearance of the IC phase in 
${\rm Pb(Zr_{1-x}Ti_x)O_3}$ is the effect of a bulk.

In this work, we have presented that the NER analysis is very effective
for systems with the KT transition and/or
with slow-dynamics which has been difficult by numerical 
methods.
Furthermore, 
a phase transition is studied using 
the finite-time scaling if we know a proper quantity
that can probe
characteristic time, $\tau$, or a characteristic length, $\xi$,
even though it is not a relevant order parameter.
It is considered that images of the phase transition which has been
believed as standard may change in some systems.

\begin{acknowledgments}
The authors would like to thank Professor Fumitaka Matsubara,
Professor Kazuo Sasaki, Professor Nobuyasu Ito and Dr. Yukiyasu Ozeki 
for fruitful discussions and comments.
They also thank Professor Nobuyasu Ito and Professor Yasumasa Kanada for
providing us with a fast random number generator RNDTIK.
\end{acknowledgments}


\begin{figure}[htb]
\epsfig{file=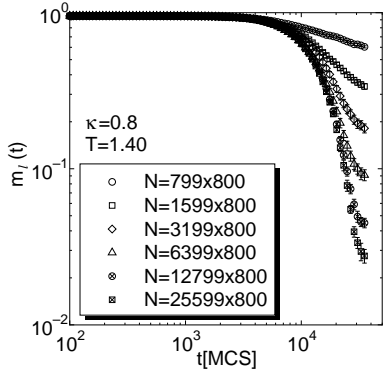,width=5.1cm}
\caption{The finite-size effect of the layer magnetization for
$\kappa=0.8$ at $T=1.4$. Length along an axial direction is fixed to
$800$, while that along a chain direction is varied from $799$ to $25599$.
The converging value is roughly considered as $\xi_x/L_x$, 
from which we derived the correlation length along the ferromagnetic 
chain direction as $\xi_x\sim500$.
}
\end{figure}

\begin{figure}[htb]
\epsfig{file=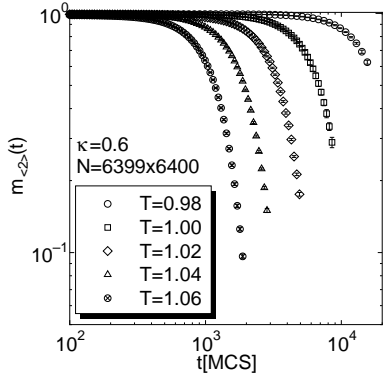,width=5.1cm}
\caption{The NER of the antiphase magnetization, 
$m_{\langle 2\rangle}(t)$, for $\kappa=0.6$.
The system size is $N=6399\times6400$.
The antiphase magnetization clearly decays exponentially at $T\ge0.98$.}
\end{figure}

\pagebreak
\begin{figure}[htb]
\epsfig{file=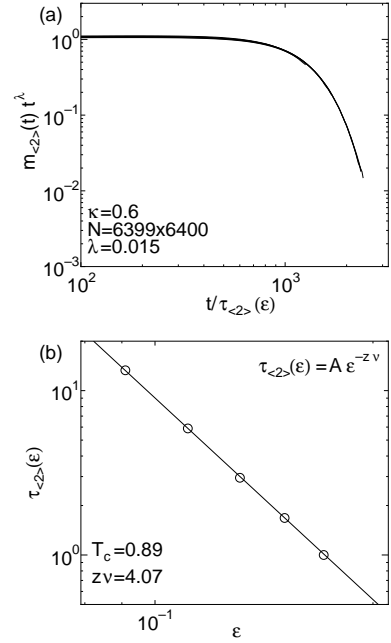,width=5.1cm}
\caption{(a)The finite-time scaling of the antiphase magnetization
 for $\kappa=0.6$, where $\lambda=0.015$.
 We use the data at five temperatures ranging $T=0.98\sim1.06$.
 (b)The least-square fitting of the relaxation time, 
 $\tau_{\langle 2\rangle}(\varepsilon)$, by Eq.~ 
 where we obtain $T_{c2}=0.89\pm0.02,z\nu=4.07$.}
\end{figure}

\begin{figure}[htb]
\epsfig{file=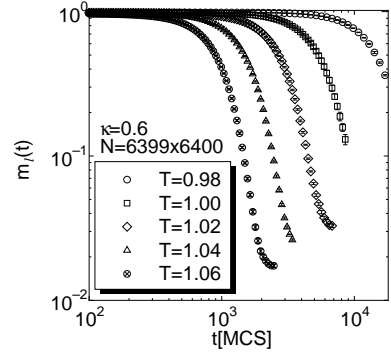,width=5.1cm}
\caption{The NER of the layer magnetization, $m_l(t)$, for $\kappa=0.6$.
 The size of system is $N=6399\times 6400$. The layer magnetization
 clearly decays exponentially at $T\ge 0.98$.}
\end{figure}

\begin{figure}[htb]
\epsfig{file=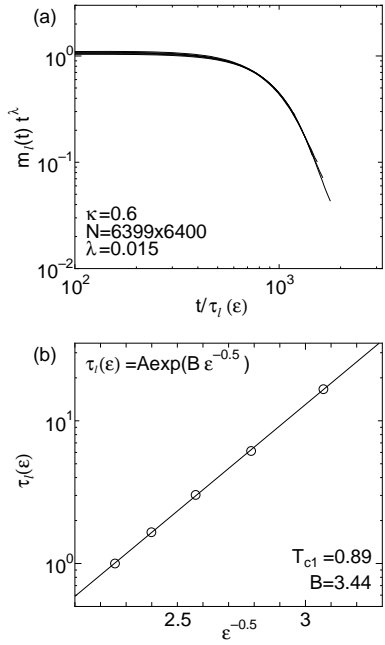,width=5.1cm}
\caption{(a)The finite-time scaling of the layer magnetization for
 $\kappa=0.6$, where $\lambda=0.015$.
 The data of five temperatures ranging $T=0.98\sim1.06$ are plotted
 together.
 (b)The least-square fitting of the relaxation time,
 $\tau_{l}(\varepsilon)$, by Eq.~(
 where we obtain $T_{c1}=0.89\pm0.02,B=3.44$.}
\end{figure}

\begin{figure}[htb]
\epsfig{file=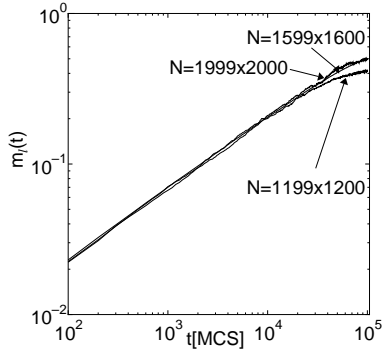,width=5.1cm}
\caption{The NER of layer magnetization, $m_l(t)$, at $T=0.92$  
 and $\kappa=0.6$. Three curves of $N=1999\times 2000$, 
 $N=1599\times 1600$ and $N=1199\times 1200$ 
 fall on the same line until $25000$ MCS.
 Then, the curve of $N=1199\times 1200$ deviates
 down from the other two curves, which remain consistent until
 $100000$ MCS.
 We discard the data of $N=1199\times 1200$ after $25000$ MCS,
 though we employ the data of $N=1599\times 1600$ until $100000$ 
 MCS.}
\end{figure}

\begin{figure}[htb]
\epsfig{file=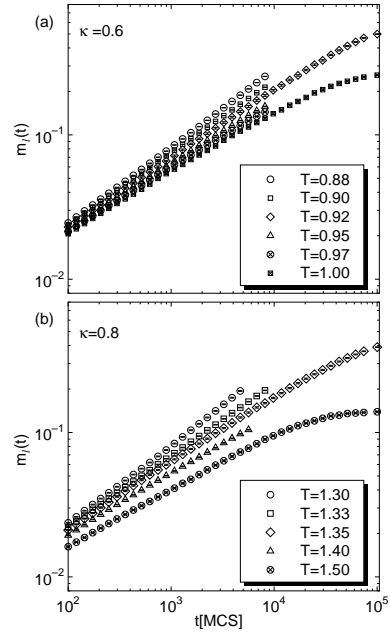,width=5.1cm}
\caption{The NER of layer magnetization, $m_l(t)$, for (a)$\kappa=0.6$
 and (b)$\kappa=0.8$ started from the paramagnetic state.
}
\end{figure}

\begin{figure}[htb]
\epsfig{file=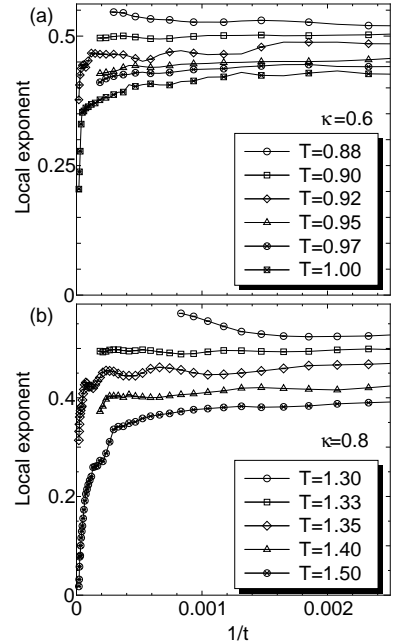,width=5.1cm}
\caption{The local exponent of the layer magnetization, $m_l(t)$, 
 for (a)$\kappa=0.6$ and (b)$\kappa=0.8$ 
 started from the paramagnetic state.
 In both systems, local exponents converge to 
 $\lambda=(2-\eta)/z=0.49$
 in the limit of $t\rightarrow\infty$ at the transition temperature.
}
\end{figure}

\begin{figure}[htb]
\epsfig{file=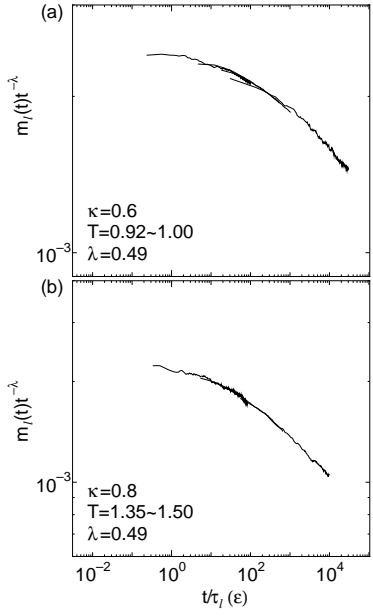,width=5.1cm}
\caption{The finite-time scaling of the layer magnetization,
 $m_l(t)$, started from the paramagnetic state.
 The data of four different temperatures ranging 
 $T=0.92\sim 1.00$ are plotted together for (a)$\kappa=0.6$,
 while those of different three temperatures ranging
 $T=1.35\sim 1.50$ are plotted for (b)$\kappa=0.8$
}
\end{figure}

\begin{figure}[htb]
\epsfig{file=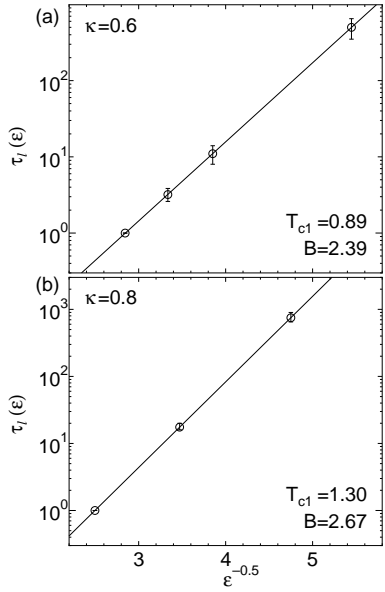,width=5.1cm}
\caption{The least-square fitting of the relaxation time, 
 $\tau_l(\varepsilon)$, by Eq.~(
 (a)$\kappa=0.6$ and (b)$\kappa=0.8$.
 The transition temperature is estimated to give the best fitting as 
 (a)$T_{c1}=0.89$ and (b)$T_{c1}=1.30$.
}
\end{figure}

\newpage
\onecolumn
\begin{table}
%
\begin{tabular}{ccccc}
 & \multicolumn{2}{@{}c@{}}{$\kappa=0.6$} &
 \multicolumn{2}{@{}c@{}}{$\kappa=0.8$}\\ 
 Present results / References (year)
 & $T_{c1}$ & $T_{c2}$ & $T_{c1}$ & $T_{c2}$\\ \hline
 NER from $T=0$ (scaling)\hspace*{4mm} & $0.89(2)$ & $0.89(2)$ 
                          & $1.31(2)$ & $1.32(2)$\\ 
 NER from $T=\infty$ (local exponent) & $0.90(2)$ & 
                                      & $1.325(25)$ & \\ 
 NER from $T=\infty$ (scaling)\hspace*{3.5mm} & $0.890(15)$ & 
                                      & $1.300(13)$ & \\ \hline
 \cite{Villain} (1981) & $\sim1.20$ & $\sim0.91$ & $\sim1.45$ &
  $\sim1.30$\\
 \cite{Selke2} (1981)  & $\sim1.40$ & $\sim1.00$ & $\sim1.70$ &
  $\sim1.50$\\
 \cite{Kroemer} (1982) & ---        & $\sim1.10$ & ---        &
  $\sim1.50$\\
 \cite{Grynberg} (1987) & $\sim1.35$ & $\sim1.05$ & $\sim1.60$ & ---\\
 \cite{Saqi} (1987)    & $\sim1.40$ & $\sim1.05$ & $\sim1.65$ &
  $\sim1.35$\\
 \cite{Murai} (1995)   & $\sim1.64$ & $0.91(1)$ & $\sim1.95$ &
  ---\\
 \cite{Sato} (1999)   & $1.16(4)$ & $\sim0.91$ & $\sim1.60$ &
  $\sim1.35$\\
\end{tabular}
\caption{Comparison the present estimate of $T_{c1}$ and $T_{c2}$ 
for $\kappa=0.6$ and $\kappa=0.8$ with the previous ones.}
\label{Tc_comparison}
\end{table}

\begin{table}
\begin{tabular}{cccccccc}
 \multicolumn{4}{c}{$\kappa=0.6$} & 
 \multicolumn{4}{c}{$\kappa=0.8$}\\
  $T$ & $\tau_{\langle 2\rangle}(\varepsilon)$ & $\tau_{l}(\varepsilon)$
  & $\tau_{l}(\varepsilon)$
  & $T$ & $\tau_{\langle 2\rangle}(\varepsilon)$ &
  $\tau_{l}(\varepsilon)$ & $\tau_{l}(\varepsilon)$\\  
  & (from $T=0$) & (from $T=0$) & (from $T=\infty$) &
  & (from $T=0$) & (from $T=0$) & (from $T=\infty$)\\
  \hline
  $1.06$ & $1.000$ & $1.000$ & ---    & $1.50$ & ---     & ---   &
  $1.000$\\
  $1.04$ & $1.680$ & $1.656$ & ---     & $1.48$ & $1.000$ & $1.000$ &
  ---\\
  $1.02$ & $2.950$ & $3.025$ & ---     & $1.46$ & $1.713$ & $1.715$ &
  ---\\
  $1.00$ & $5.889$ & $6.145$ & $1.000$ & $1.44$ & $3.033$ & $3.140$ &
  ---\\
  $0.98$ & $13.301$ & $16.657$ & ---     & $1.42$ & $6.264$ & $6.889$ &
  ---\\ 
  $0.97$ & ---     & ---     & $3.202$ & $1.40$ & $14.822$ & $20.190$ &
  $17.503$\\
  $0.95$ & ---     & ---     & $11.103$ & $1.35$ & ---     & ---     &
  $752.314$\\
  $0.92$ & ---     & ---     & $500.115$ & & & &\\
\end{tabular}
\caption{The relation between the temperatures, $T$, and the relaxation time,
 $\tau_{\langle 2\rangle}(\varepsilon)$, of the antiphase magnetization, 
 $m_{\langle 2\rangle}(t)$, and the relaxation time,
 $\tau_{l}(\varepsilon)$, of the layer magnetization, $m_{l}(t)$, 
 for $\kappa=0.6$ and $0.8$.
 The relaxation time is scaled so that a value at the highest
 temperature becomes unity.}
\label{tau_0.6}
\end{table}

\begin{table}[htb]
\begin{tabular}{cccc}
  & $T$ & $L_{x}\times L_{y}$ & Crossover MCS\\ 
\hline
  $\kappa=0.6$ & $0.88$ & $999\times1000$ & $\sim 8000$ \\ 
  & $0.90$ & $999\times1000$ & $\sim 10000$ \\ 
  & $0.92$ & $1599\times1600$ & $\sim 100000$ \\ 
  & $0.95$ & $999\times1000$ & $\sim 10000$ \\ 
  & $0.97$ & $799\times800$  & $\sim 8000$ \\  
  & $1.00$ & $999\times1000$  & $\sim 100000$ \\ \hline
  $\kappa=0.8$ & $1.30$ & $799\times800$ & $\sim 5000$ \\
  & $1.33$ & $999\times1000$ & $\sim 10000$ \\
  & $1.35$ & $1199\times1200$  & $\sim 100000$  \\
  & $1.40$ & $599\times600$  & $\sim 8000$  \\
  & $1.50$ & $599\times600$  & $\sim 100000$  \\
\end{tabular}
\caption{The crossover MCS at each temperature for $\kappa=0.6$ and
 $0.8$.
 This is a MC step until which the NER function falls on the same line
 as that of a larger system.}
\label{size_effect}
\end{table}

\end{document}